\documentclass{SCGE}
\usepackage{multicol}
\usepackage{epsfig}
\begin{document}

\begin{picture}(0,0){\rm
\put(0,-39){\makebox[160truemm][l]{\bf {\sanhao\raisebox{2pt}{ }}
Review Paper
 {\sanhao\raisebox{1.5pt}{ }}}}}
\end{picture}

\def\bm{\boldsymbol}

\def\dl{\displaystyle}
\def\du{\end{document}}

\Year{2013} %
\Month{July}
\Vol{56} %
\No{7} %
\BeginPage{1} %
\EndPage{12} %
\AuthorMark{{\rm CHEN SongZhan,} }
\DOI{10.1007/s11433-013-5128-z} 

\title{Observations of very high energy gamma-ray  emission from AGNs with the ground-based EAS arrays}

\author[1*]{CHEN SongZhan}{}
\address[{\rm1}]{Key Laboratory of Particle Astrophysics, Institute of High Energy Physics, Chinese Academy of Sciences, Beijing 100049, China;}

\maketitle \vspace{-3.5mm}{\footnotesize\begin{center} Received December 3, 2012; accepted January 7, 2013
\end{center}}\vspace*{-5mm}

\begin{center}
\rule{16.5cm}{0.4pt}
\parbox{16.5cm}
{\begin{abstract}
 The  ground-based EAS array is usually operated with a high duty cycle ($>$ 90\%) and a large field of view ($\sim2 sr$), which can continuously monitor the  sky. It is  essential and irreplaceable to understand the  gamma-ray emission mechanism and intrinsic physics progress of  the variable source AGN. The EAS arrays, AS-$\gamma$ experiment (since 1990) and ARGO-YBJ experiment (since 2007),
have  continuously monitored the northern sky at energies above 3 TeV and 0.3 TeV, respectively. They have made substantial contributions for long-term monitoring of Mrk 421 and Mrk 501. In this paper, we will review the results obtained by the EAS arrays. The next generation of EAS array, LHAASO project, will boost the sensitivity of current EAS array at least up to 30 times with a much wider energy range from 40 GeV to 1 PeV. Beside increasing the number of VHE gamma-ray sources, it will  guide us look sight into the properties of jet, and  throw light on the determining of the EBL, intergalactic magnetic fields, and the validity of the Lorentz Invariance.

\end{abstract}}
\end{center}\vspace*{-0.6cm}
\begin{center}
\parbox{16.5cm}{\bf\jiuhao active galactic nuclei(AGN), gamma-ray, non-thermal, extensive air shower (EAS) array  
}
\end{center}

\begin{center}
\parbox{16.5cm}{\PACS{\hspace*{-2mm}\rm 95.55.Vj, 95.85.Pw, 95.85.Ry, 98.54.Gr, 98.70.Rz}
\rule{16.5cm}{0.4pt}}\end{center}



\wuhao\vspace*{1.5mm}
\begin{multicols}{2}
\renewcommand{\baselinestretch}{1.08} \baselineskip 12.2pt\parindent=10.8pt

\no 

\section{Introduction}
Very high energy (VHE;E$>100$GeV) gamma-ray astronomy is  a new  window to explore the extreme non-thermal phenomena in the universe.
VHE gamma-rays are tracers of non-thermal particle acceleration and are  used to probe the condition and underlying
astrophysical processes inside the sources.
In the past decade, more than one hundred VHE gamma-ray emitter were observed, and about 40\% were active galactic nucleis (AGNs).
An  AGN is a compact region at the center of a galaxy that has a much higher than normal luminosity over all of the electromagnetic spectrum. The radiation from AGN is believed to be a result of accretion of mass by the super-massive black hole at the center of the host galaxy. Blazars, including BL Lac objects and flat-spectrum radio quasars (FSRQs), are the most extreme subclass of AGNs known, and most of the identified extragalactic gamma-ray sources detected with the $Fermi$-LAT instrument [1] and ground-based VHE gamma-ray detectors [2] belong to this category. The  emission is believed to be dominated by
\\
\noindent\rule{2.5cm}{0.4pt}\\[0.1mm]{\qihao *Corresponding author (email:
chensz@ihep.ac.cn)}
\\
 non-thermal and strongly Doppler-boosted radiation from
a relativistic jet of magnetized plasma which is aligned along our line of sight, while the production of the jet  is still  unknown. Currently, even though about a thousand of GeV gamma-ray blazars [1] and about 50 VHE gamma-ray blazars [2] have been observed, the physical mechanisms responsible for the production of their gamma-ray emission has not been unambiguously identified.

The spectral energy distribution (SED) from blazar is double-humped.
The low energy hump  from the radio to X-ray bands is usually
interpreted as  synchrotron radiation from relativistic
electrons (and positrons) within the jet. The origin  of the
high energy hump at GeV-TeV gamma-ray is under debate. The leptonic models attribute it to the inverse
Compton scattering of the synchrotron (synchrotron
self-Compton, SSC) or external photons (external Compton,
EC) by the same population of relativistic electrons [3,4], therefore a X-ray/gamma-ray correlation would be expected. The hadronic models invoke hadronic processes including proton-initiated cascades and/or proton-synchrotron emission in a magnetic field-dominated jet [5]. The emission from AGN are variable with timescales from sub-hour to months, particularly at X-ray and VHE gamma-ray bands. Recently, an important  progress is that rapid flares up to  minutes have been detected  from   AGNs [6]. The luminosity can be enhanced up to tens or even hundreds times brighter than usual during the flares. It is unclear what causes these   variation.  The  high variability and
broadband emission require  long-term, well-sampled, multiwavelength
observations in order to understand the radiation mechanisms.

The low energy hump is well sampled during the multi-wavelength campaigns, including radio, optical, X-ray. The high energy hump has been continuously monitoring by $Fermi$-LAT from 100 MeV to  300 GeV since August 2008.  Because of  fast decreasing of the emission flux, the observation of the gamma-ray at  VHE band  can only be achieved by ground-based detectors with an indirect way, which can detect the
secondary particles of the atmospheric showers produced by
the primary photons. The  effective collection areas are usually larger than 10 thousand square meters.
 There are two main classes of ground-based
detector techniques : the Extensive Air Shower array (EAS) and the Imaging Atmospheric Cherenkov Telescope (IACT).

In this paper, we will present the   progress of ground-based EAS arrays for AGN observation. The results about AGN obtained by EAS arrays are reviewed as well as a brief   discussion of  the future EAS array,  using LHAASO as an example for the    AGN and its relative field study.

\section{The progress of ground-based EAS arrays}
In the past decade,  the IACT technique has provided sufficient data.  The current IACTs,  such as  H.E.S.S., VERITAS and MAGIC, have  excellent  angular resolution $\sim$0.1$^{\circ}$ and powerful gamma-ray/hadron discrimination, which together lead to a high sensitivity $\sim$1\% Crab flux in 50 hours observation.
All the known AGNs with VHE emission and their important general features  are discovered by IACTs, which have made  substantial contributions   to our knowledge about the structure and composition of the highly relativistic jet from AGN.
As the most sensitive VHE gamma-ray detectors, however, IACTs have extra technical difficulties to constantly monitor
AGNs because their narrow field of view (FOV, $3^{\circ}\sim5^{\circ}$) and limited duty cycle($\sim$10\%). The practical observation time for a specific AGN is much less, such as the duty cycle to the most frequently sampled AGN Mrk421 is less than 0.1\% even all the worldwide IACT data are combined [7],   which constrains the systematic study on the long-term temporal feature of the variable sources.
A long-term observation  is better performed   by the wide FOV
EAS arrays, such as the Tibet AS-$\gamma$ experiment  and ARGO-YBJ experiment, which are operated   continuously with a duty cycle higher than 90\% and can observe any source with a zenith angle less than 50$^{\circ}$. The practical duty cycle for any specific AGN in the sky is about 20\%.

Comparing to the success of IACTs,
the development of EAS arrays proceeded more slowly.  The first generation of EAS arrays  adopted sample detector technique. The typical experiment is the Tibet AS-$\gamma$ array, which is located in Yangbajing, Tibet,  China, has  an altitude of 4300 m above sea level (a.s.l.). It was constructed in 1990 using plastic scintillation detectors  with a coverage of 0.25\%. This array was gradually expanded and the final coverage is improved to be 1\%. Schematic view of the Tibet AS-$\gamma$ array is shown in  Fig.1. Among many EAS arrays, only Tibet AS-$\gamma$ array successfully detected one AGN Mrk 421 with significance of 5 $\sigma$  [8], demonstrating the sensitivity of the EAS array technique for AGN study.  Moreover,  Tibet AS-$\gamma$ array also detected the Mrk 501 with a marginal significance during the  huge flare in 1997 [9].  Currently, a 10,000 m$^2$ water-Cherenkov-type muon detector (MD) array is  planned to be constructed  under the Tibet AS-$\gamma$ array, which will improve its sensitivity to be about 10\% Crab flux   at energy from 10 TeV to 100 TeV.

\vspace*{8mm}
\begin{center}
\centerline{\psfig{figure=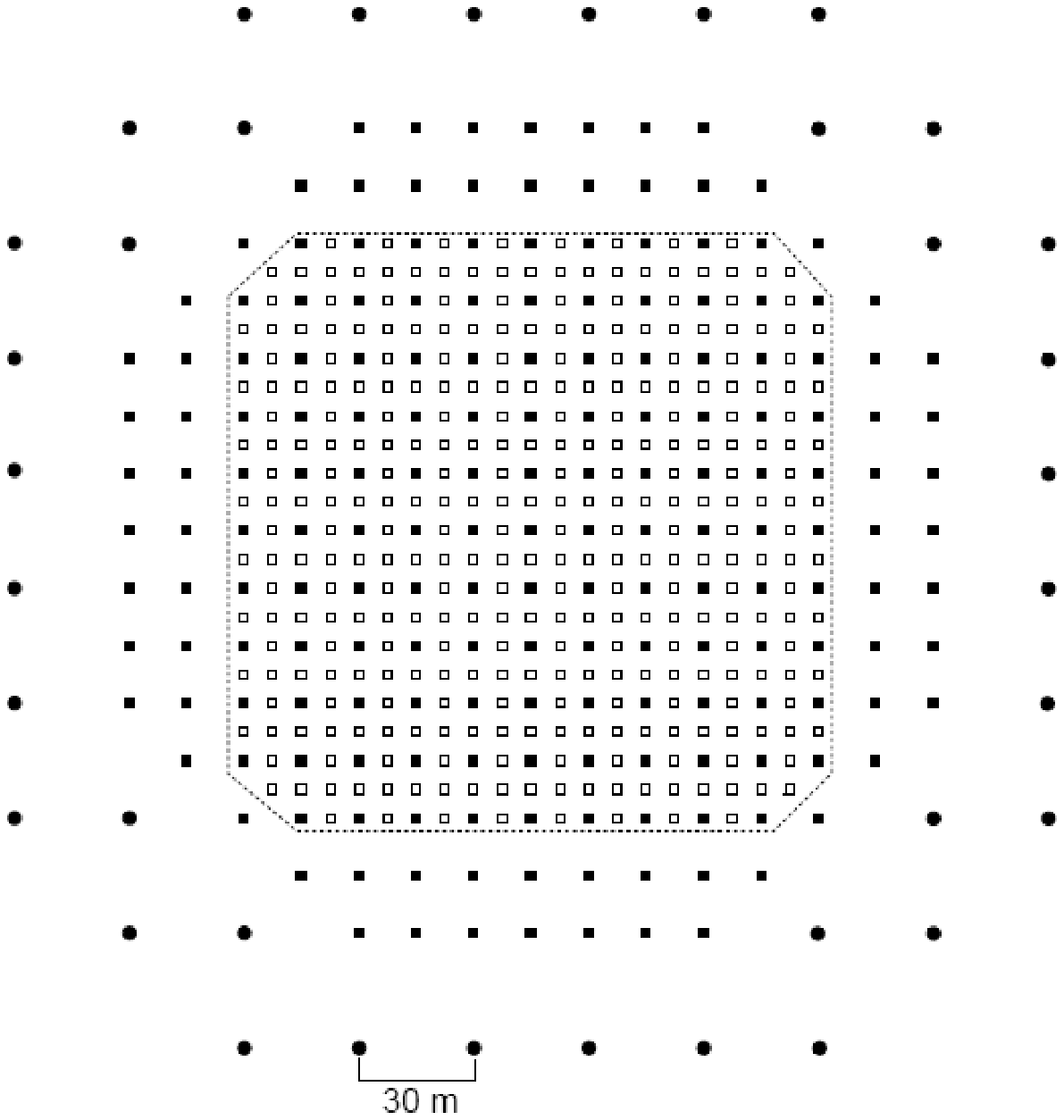,width=3.in,height=2.8in}}
\end{center}
\vspace*{-5mm} {\footnotesize {\bf Figure 1}\quad Schematic view  of the
Tibet AS-$\gamma$ array at Yangbajing [8]. Reproduced by permission of the AAS.}

 To reduce the threshold, a full coverage detector technique is adopted by the second generation EAS arrays, such as Milagro and ARGO-YBJ experiment. The Milagro detector (1999$\sim$2008) ,  located in  Arizona, USA,  at an altitude of 2630 m  a.s.l., use water Cherenkov detector within a 80 m$\times$60 m$\times$8 m water pool. Milagro has detected the AGN Mrk 421 using several-year data [10], while its sensitive energy is shifted to above  20 TeV after using new analysis method, which is not optimized for AGN observation.   The  ARGO-YBJ located in Tibet  an altitude of 4300 m a.s.l. consists of  a full coverage of resistive plate chambers [11]. A schematic view of the detector is shown in  Fig.2. Since  2007 ARGO-YBJ has been in stable operation with average duty cycle higher than 86\%.  With a better sensitivity and much lower threshold than any other EAS array, ARGO-YBJ
can detect both the long-term steady emission and several-day flares from the Mrk 421 [12,13] and Mrk 501 [14]. Combining the   X-ray and GeV data obtained by satellite-borne detectors, a long-term continuous multi-wavelength investigation has been naturally  achieved by the ARGO-YBJ collaboration [12], which
has provided import temporal and spectral evidence for the emission from AGN jet and demonstrated the essential and irreplaceable role of EAS array in monitoring the VHE emission from AGN [12].

\vspace*{8mm}
\begin{center}
\centerline{\psfig{figure=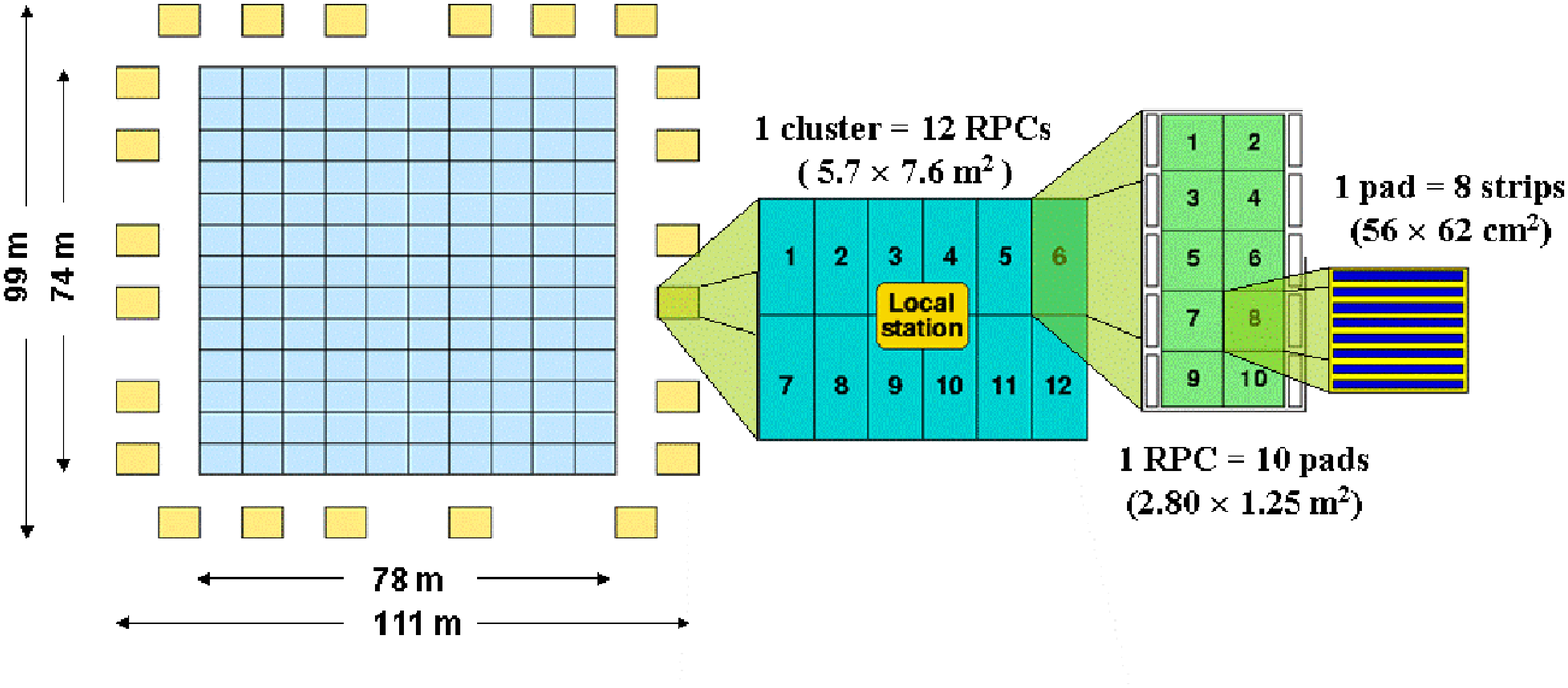,width=3.3in,height=1.6in}}
\end{center}
\vspace*{-5mm} {\footnotesize {\bf Figure 2}\quad Schematic drawing of the ARGO-YBJ experimental setup, with the details of
the structure of one cluster and one pad [11].   }

\vspace*{8mm}
\begin{center}
\centerline{\psfig{figure=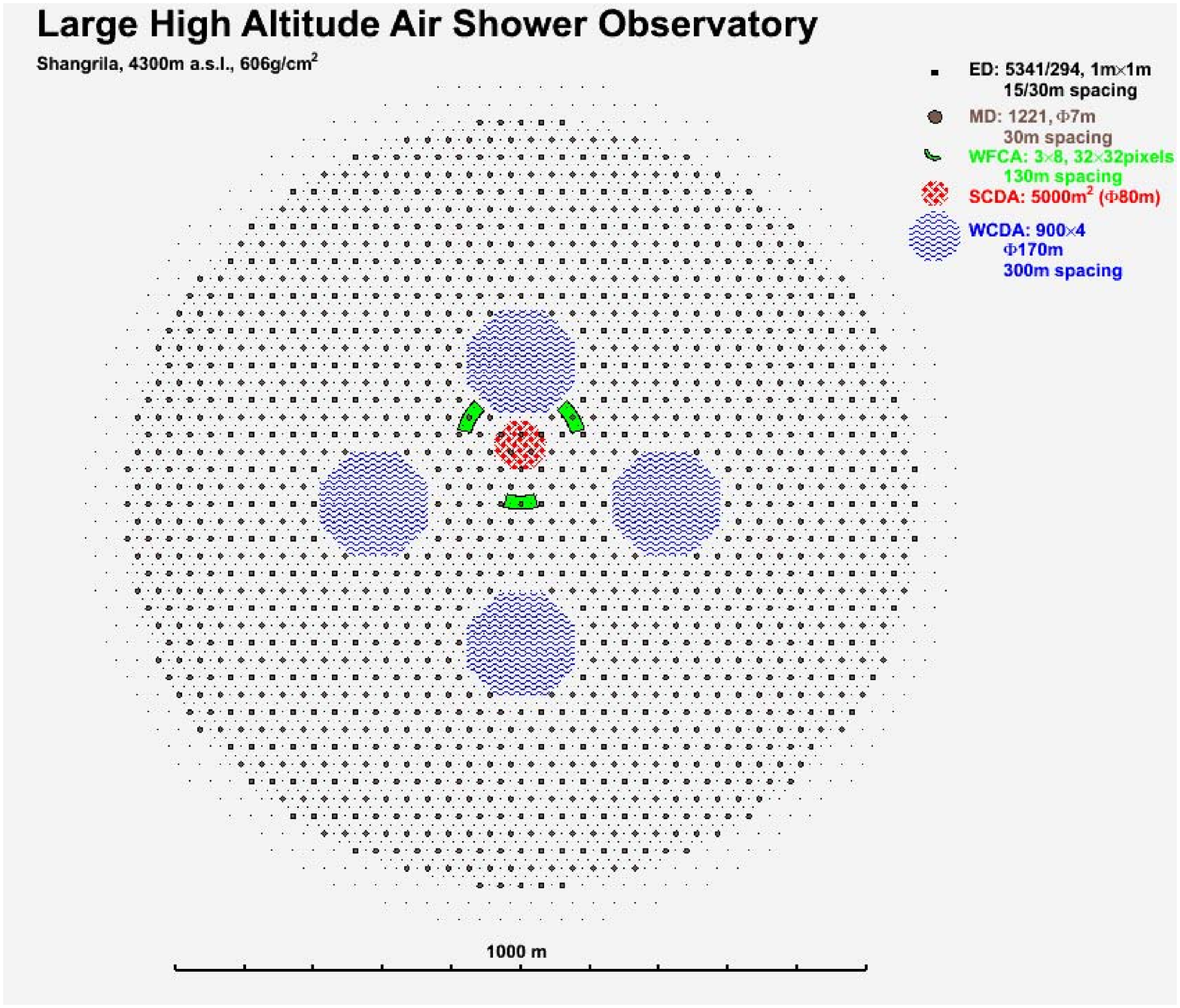,width=3.3in,height=2.8in}}
\end{center}
\vspace*{-5mm} {\footnotesize {\bf Figure 3}\quad A layout of the LHAASO array [16].  }

Fortunately,  the next generation of EAS arrays HAWC and LHAASO projects  will boost the sensitivity to be at least 10 times higher than  current EAS array ARGO-YBJ. HAWC will use the water Cherenkov technique and  be located at altitude of 4100 m a.s.l. in the Sierra Negra, Mexico [15]. Such high altitude will significantly lower the threshold. Its sensitive area will  be 22,500 m$^2$, which is about 5 times larger than its predecessor Milagro.  The  LHAASO [16] is a large EAS array with   multi-techniques in an area of 1 km$^2$ and will be located in Shangrila, China at  an  altitude of 4300 m a.s.l.. For gamma-ray sources surveys, this project will include a water cherenkov array (WCDA) with a similar design as HAWC while the sensitive area is 4 times larger, 90,000 m$^2$. To extend energy up to hundreds of TeV for gamma-rays, this project will also include   a particle detector array with an effective area of 1 km$^2$ (KM2A) including a muon detector array with 40,000 m$^2$ active area. A schematic view of the detector is shown in  Fig.3.  The LHAASO project will boost the sensitivity of current EAS array at least up to 30 times with a much wider energy range from 40 GeV to 1 PeV.

\vspace*{4mm}
\begin{center}
\centerline{\psfig{figure=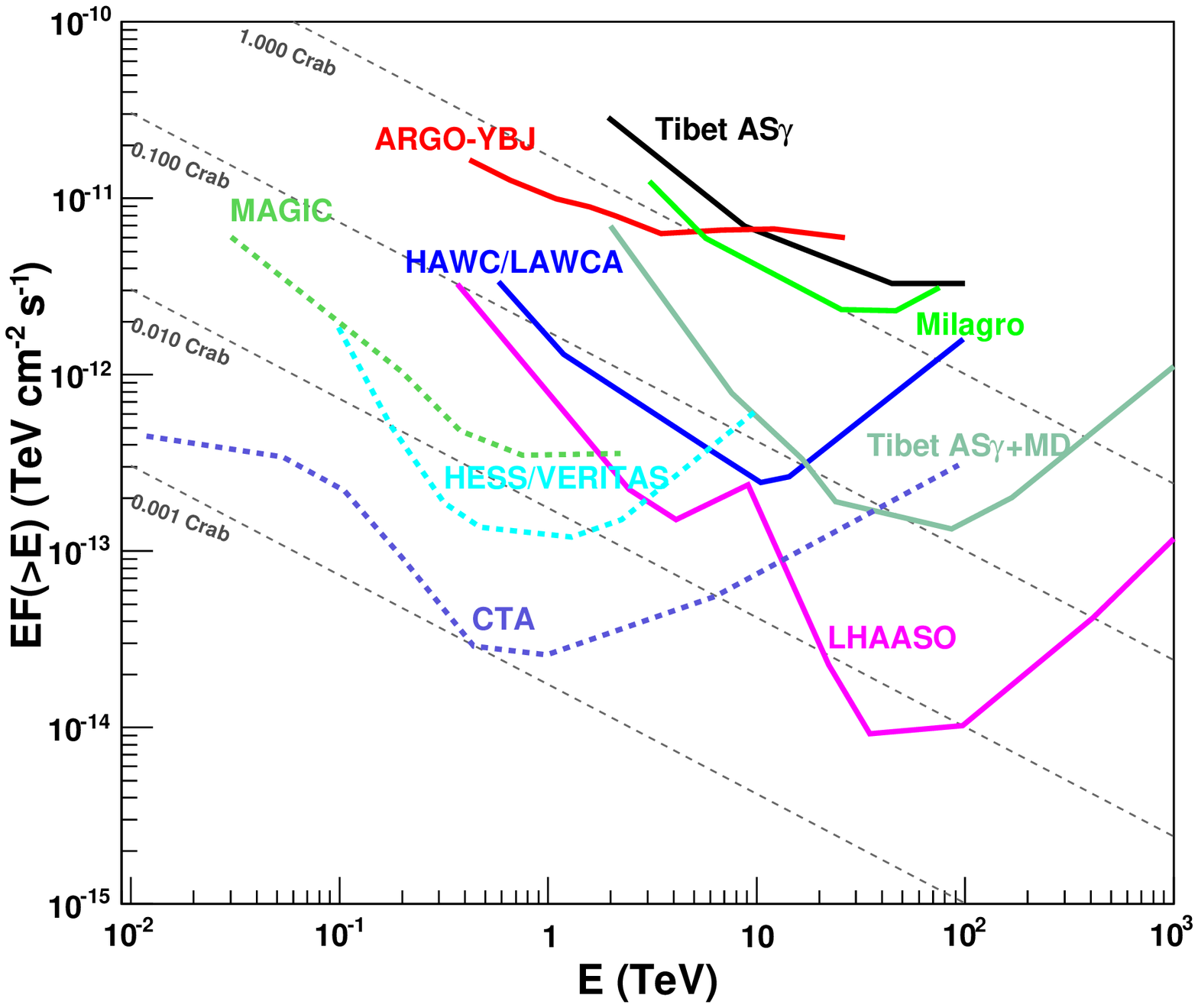,width=3.3in,height=2.5in}}
\end{center}
\vspace*{-5mm} {\footnotesize {\bf Figure 4}\quad Sensitivity of the major experiments and future projects for gamma-ray astronomy [16].   The observation time is 1 y and
50 h for EAS arrays (solid lines) and IACTs (dotted lines), respectively.  }

The progress of ground-based EAS arrays can be clearly seen from the improvement of sensitivity shown in
 Fig.4. In one year observation, the detectable source flux is higher than 1 Crab flux for the first generation EAS array Tibet AS-$\gamma$ experiment, about 0.5 Crab flux for the second generation EAS array ARGO-YBJ experiment, and about 2\% Crab flux for the next generation EAS array LHAASO project.
For  comparison, the sensitivities for current IACTs and future IACT, that is  CTA, are also shown in  Fig.4. The observation time for IACTs is 50 hour.

\section{AGN long-term monitoring with EAS arrays}
 Limited  by the sensitivity, only two AGNs, Mrk 421 and Mrk 501, were detected by the first and second generations of EAS arrays.
However, the EAS arrays are   operated with a high duty cycle ($>$ 90\%) and a large field of view ($\sim2 sr$), which   have made substantial contributions for long-term monitoring of Mrk 421 and Mrk 501. Both Tibet AS-$\gamma$ and ARGO-YBJ experiments have reported their specific observations for Mrk 421 and Mrk 501 in the past decade. We will review the observation findings for Mrk 421 and Mrk 501, respectively, in the following.

\subsection{Mrk421}
Mrk 421 ($z=0.031$) is one of the brightest blazars known. It was  the first extragalactic object detected by a ground-based experiment (Whipple) at energies around 1 TeV [17]. Mrk 421 is a very active blazar with   major outbursts about once every two years in  X-ray, as shown in  Fig.5. The light curves from Mrk 421 at energy    (2$-$12 keV) is provided by   ASM/$RXTE$\footnote{Quick-look results provided by the ASM/$RXTE$ team:
\url{http://xte.mit.edu/ASM_lc.html}.}.
 A major outburst usually lasts
several months and is accompanied by many rapid flares with timescales from tens of minutes to several days.
Mrk 421 is  currently the most studied  blazar at VHE gamma-ray band with extensive data on various timescales, as shown in  Fig.5.
However, due to the limitation of the IACT technique as discussed in Section 2,
there were no simultaneous IACT observations for most of the  flares presented in X-ray band. The observations for most of the outbursts were also not complete, as shown in  Fig.5, which is difficult to  infer the physics of the flare production and decay. Therefore, even though some important general features of the AGN flares have been obtained,
most of the results still suffer from various possible experimental caveats [18].

\vspace*{2mm}
\begin{center}
\centerline{\psfig{figure=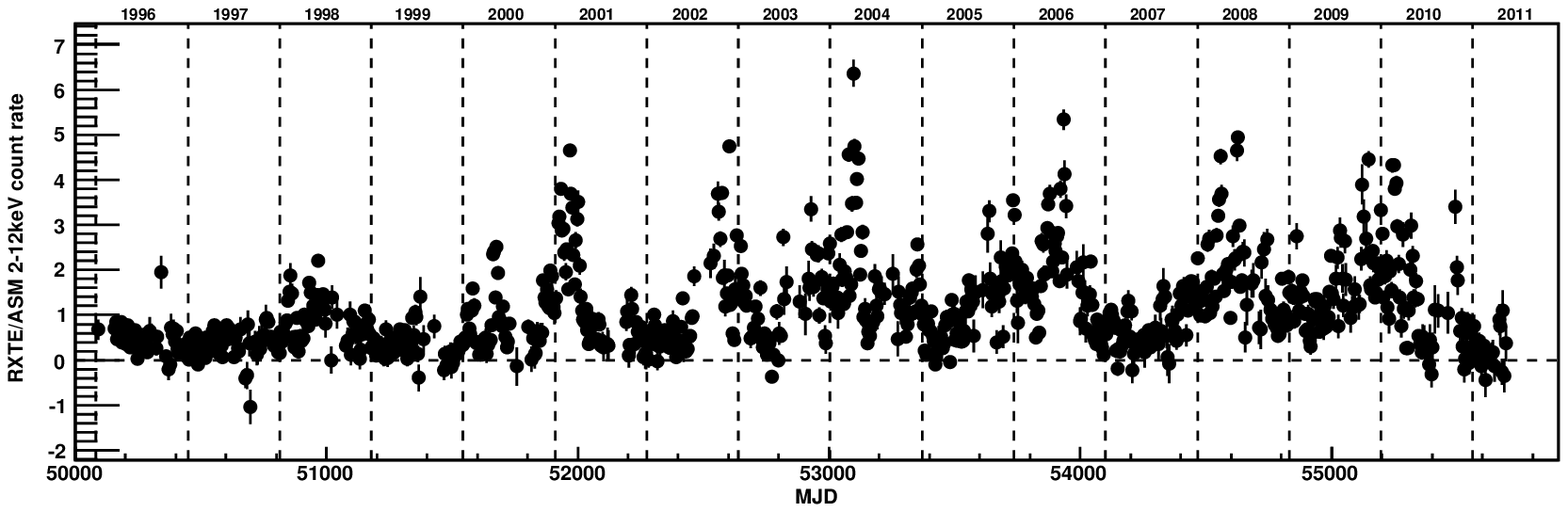,width=4.0in,height=1.5in}}
\centerline{\psfig{figure=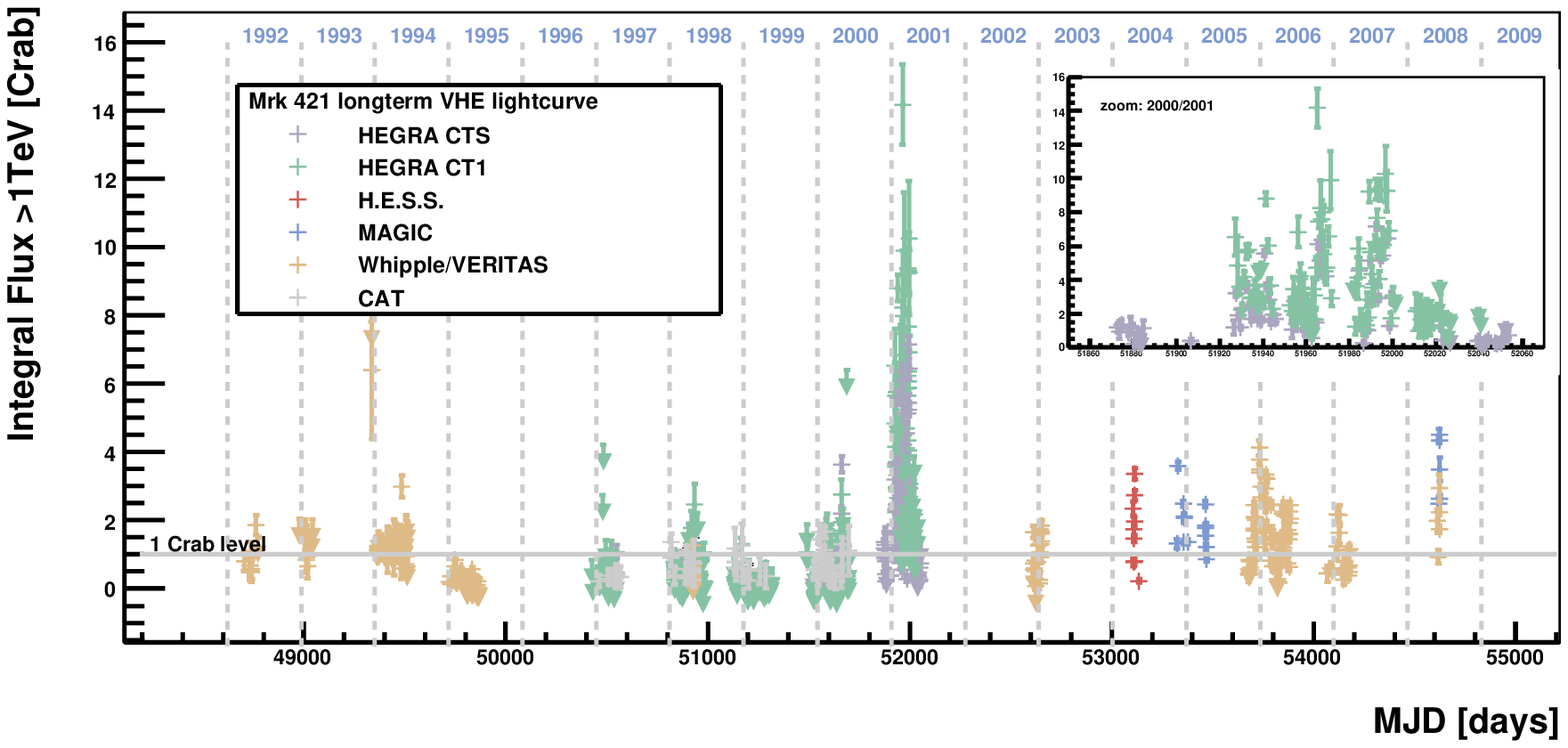,width=3.3in,height=1.5in}}
\end{center}
\vspace*{-3mm} {\footnotesize {\bf Figure 5}\quad Long-term light curve of Mrk 421. Up: Light curve observed by RXTE/ASM detector at 2-12keV.
Each bin contains the event rate averaged over the
five-day interval centered on that bin. Down:  A combined long-term light curve using data from  the major VHE gamma-ray telescopes, which is presented in [7].
 Reproduced by permission of Astronomy and Astrophysics. }

During this flaring period in 2000 and 2001, Mrk 421 is observed by Tibet AS-$\gamma$  at a significance
level of 5.1 $\sigma$ [8].  A positive flux correlation between the keV and TeV energy regions is finding. It should be stressed that this is the first observation of long-term correlations between satellite keV X-ray and TeV gamma-ray data based on simultaneous observations.
The light curves from X-ray, Tibet AS-$\gamma$ and IACTS are shown in  Fig.6.
\vspace*{2mm}
\begin{center}
\centerline{\psfig{figure=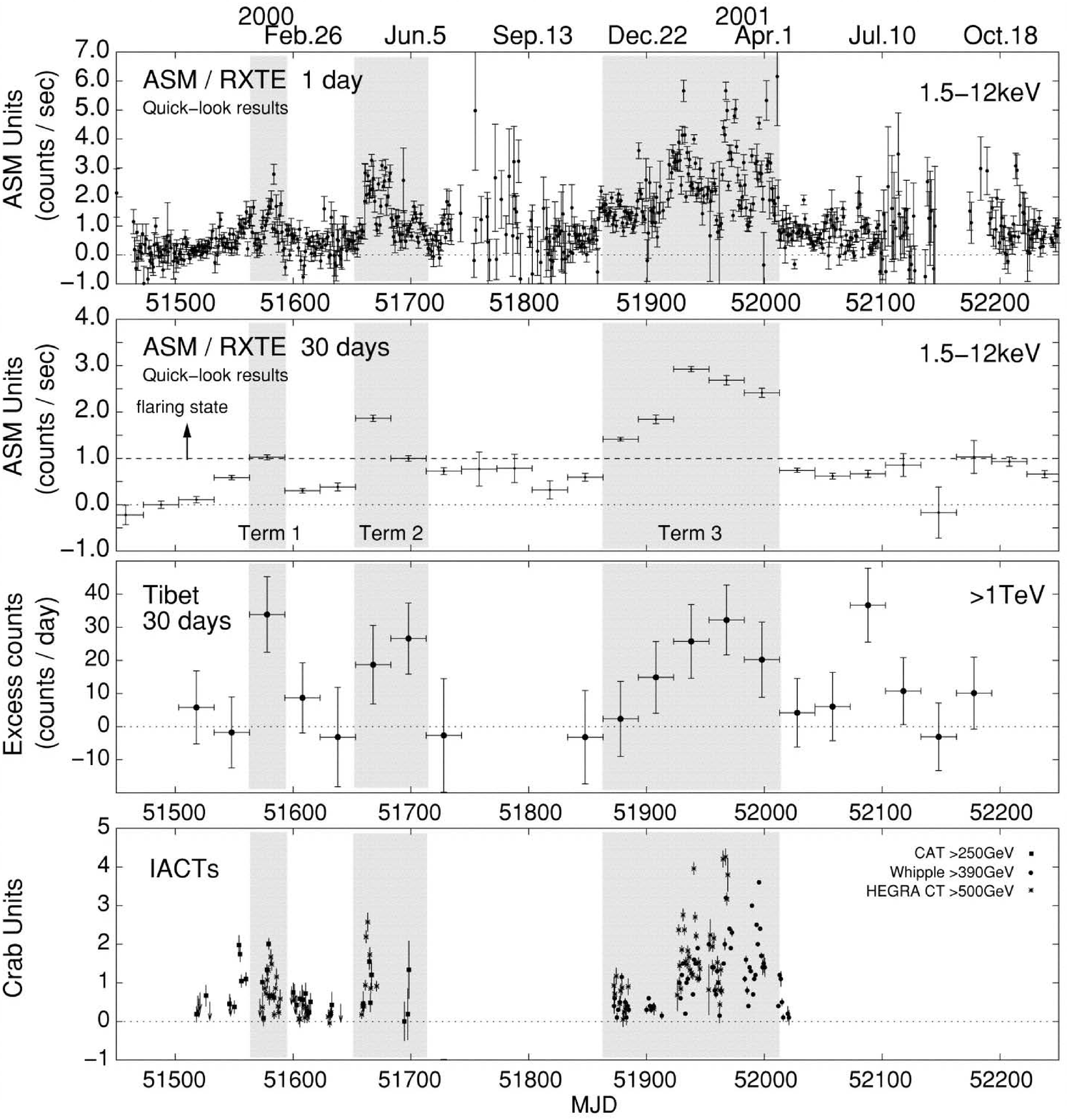,width=3.3in,height=3in}}
\end{center}
\vspace*{-3mm} {\footnotesize {\bf Figure 6}\quad
Light curves are from X-ray, Tibet AS-$\gamma$ and IACTs around the Mrk 421 flaring period in the years 2000 and 2001.
The figure is taken from [8]. Reproduced by permission of the AAS. }

 \vspace*{2mm}
\begin{center}
\centerline{\psfig{figure=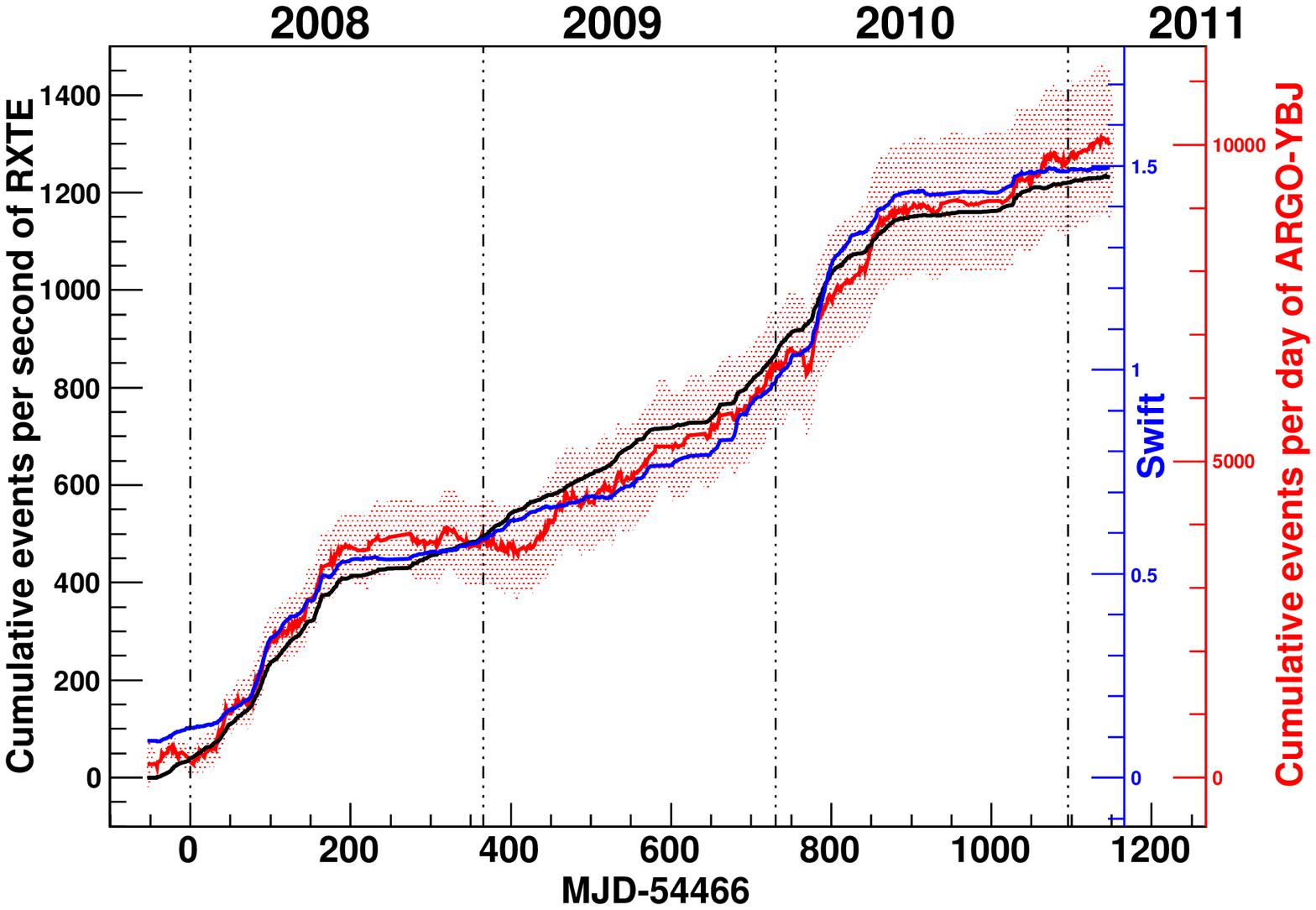,width=3.6in,height=2in}}
\end{center}
\vspace*{-3mm} {\footnotesize {\bf Figure 7}\quad Cumulative
light curves from the Mrk 421 direction:  the gamma-rays are   observed by ARGO-YBJ,
soft X-rays (2-12 keV) are observed by ASM/RXTE; hard X-rays (15-50 keV)
 are observed by BAT/Swift. The figure is taken from [20].  }

Mrk 421 has been continuously monitored by ARGO-YBJ for 4 years since  2007   at energy above 0.3 TeV. The flux has been detected by ARGO-YBJ with significance greater than 12$\sigma$.
Combining the ARGO-YBJ data  with the X-ray data, a long-term simultaneous multiwavelength observation was achieved [12].
 Fig.7 shows the accumulation rates of Mrk 421 events detected by ARGO-YBJ, Swift and RXTE, which clearly reveal the long-term variation.
A clear correlation is visible among the three curves. The rapid increases of the rates during the first half of 2008 and 2010 are related to the strongest active periods, when the flux increased up to the level of several Crab units. During the outburst of 2008, four large flares are observed by
all three detectors, and the peak times are in good agreement with one another, as shown in  Fig.8. It should be stressed that the observation for the fourth flare was important when the IACTs were hampered by the Moon [13].
During the outburst of 2010, the strong flare on February
16, 2010 is detected in all four energy bands from soft X-ray to VHE gamma-ray and their time of peak flux are consistent with each other [19].

\vspace*{2mm}
\begin{center}
\centerline{\psfig{figure=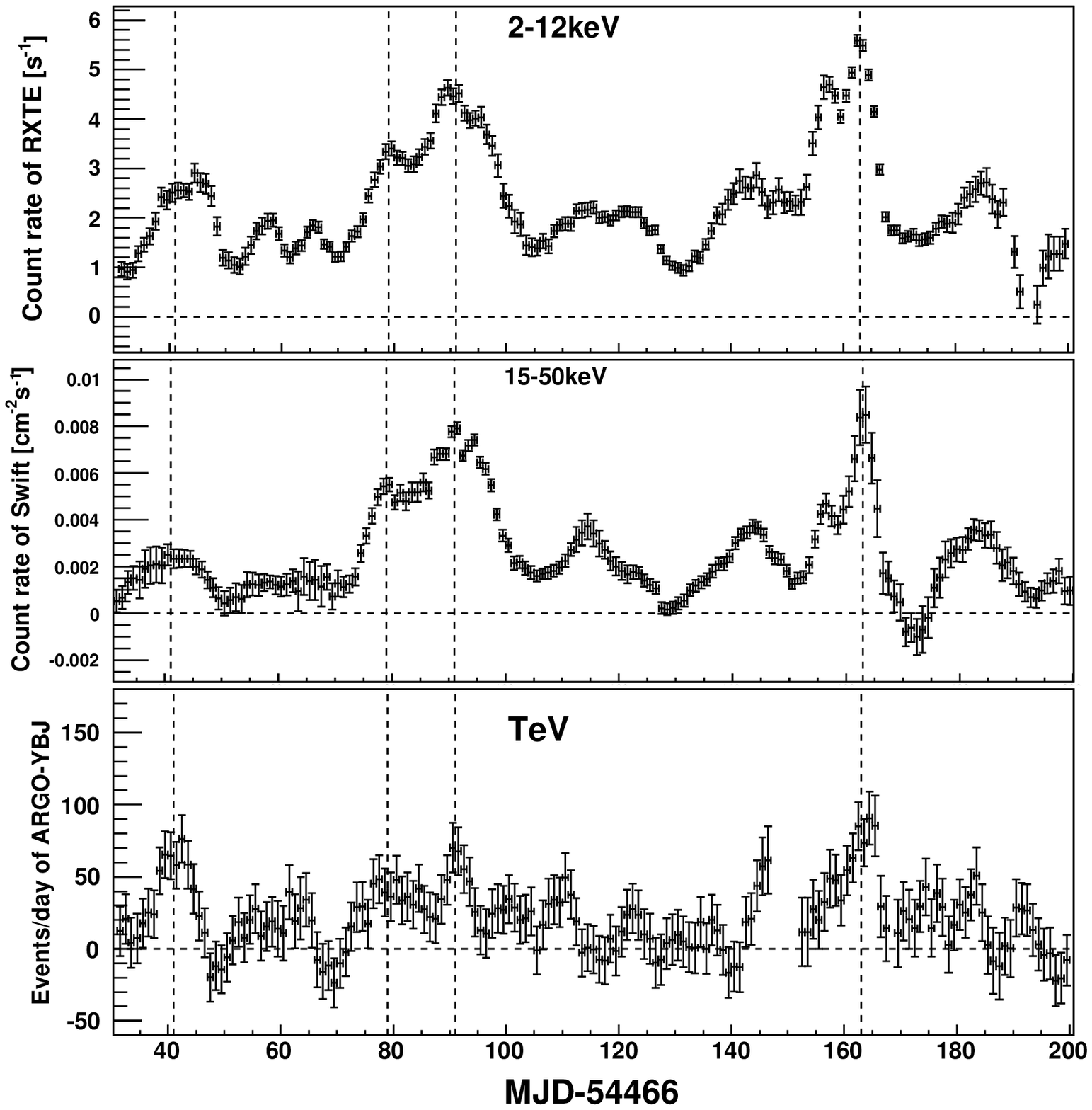,width=3.6in,height=2.5in}}
\end{center}
\vspace*{-3mm} {\footnotesize {\bf Figure 8}\quad Daily light
curves from Mrk 421 direction in different energy bands from  2008 February 1 to July 18. Each bin contains the event rate averaged over the five-day interval centered on that bin. The panels from top to bottom refer to 2$-$12 keV (ASM/$RXTE$),
 15$-$50 keV (BAT/$Swift$), and gamma-ray (ARGO-YBJ),
respectively. The figure is taken from [12]. Reproduced by permission of the AAS.}

\vspace*{2mm}
\begin{center}
\centerline{\psfig{figure=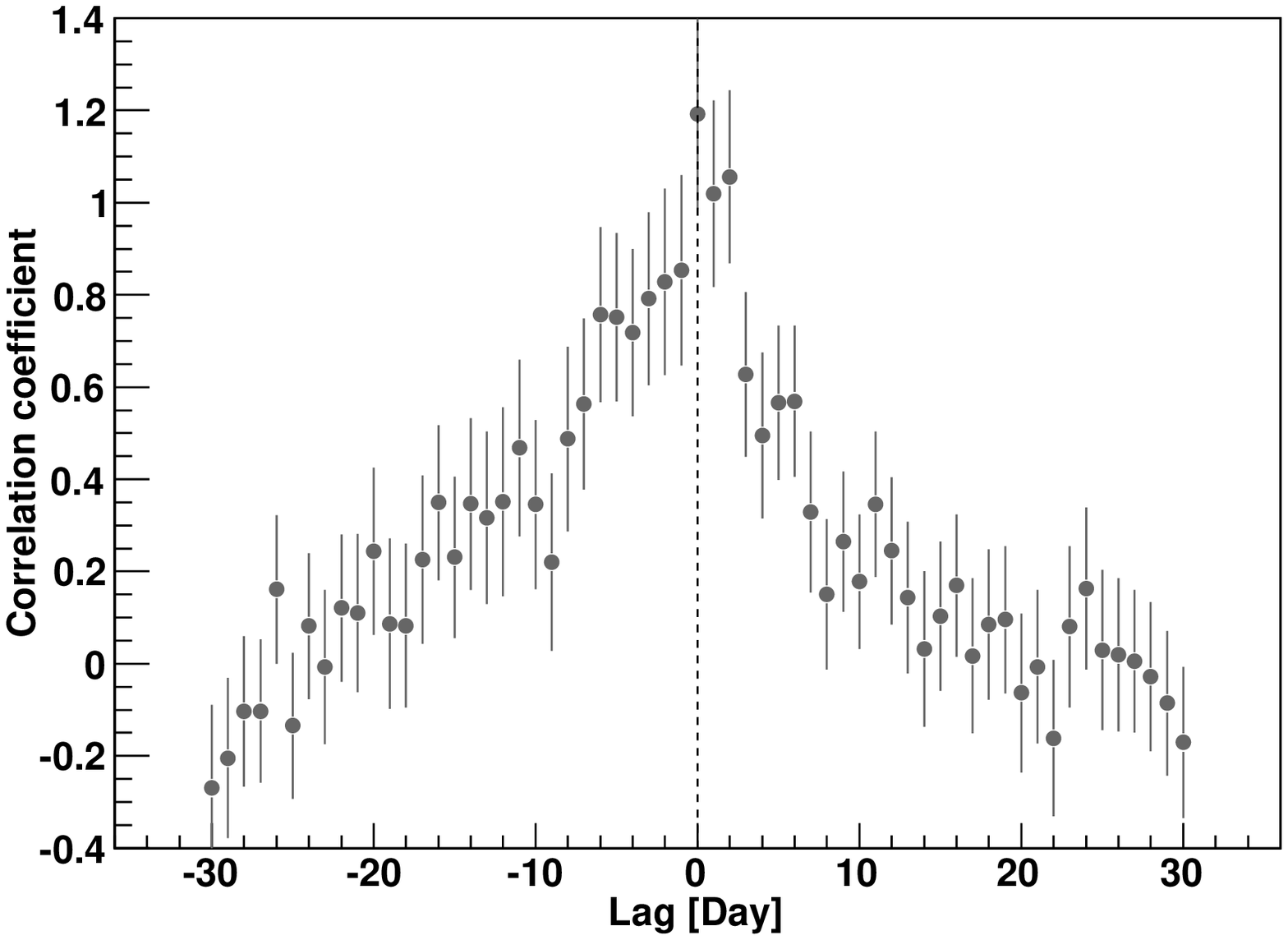,width=3.2in,height=2in}}
\end{center}
\vspace*{-3mm} {\footnotesize {\bf Figure 9 }\quad Discrete correlation function between  X-ray (2-12 keV) and gamma-ray light curves. The figure is taken from [20].}

The  long-term simultaneous multiwavelength observation   covers both active and quiet phases without experimental caveats, which is crucial to
quantify the degree of correlation and the phase differences (lags) in the variations between gamma-rays and X-rays.
The  discrete correlation function (DCF)  derived from RXTE and ARGO-YBJ data is shown in  Fig.9  with the peak of the distribution around zero. No significant lag longer than 1
day is found. This result is contrast with that obtained in [21] that a lag of about
two days between X-rays and gamma-rays, which  suffered  from experimental caveats.

Both the Tibet AS-$\gamma$ and ARGO-YBJ results presented in  Fig.6-9  show a good long-term correlation between X-ray flux and the VHE gamma-ray flux, indicating that the emission at the two energy bands have a common origin as assumed in the SSC model.
The evolution of the spectral energy distribution is investigated by measuring spectral indices at four
different flux levels [12]. Hardening of the spectra is observed in both X-ray and gamma-ray bands, which are also expected in the SSC model.

In the framework of SSC mechanism, the flux of synchrotron photons is proportional to the electron density, and the flux of IC VHE gamma-ray flux is  proportional to both the  electron density and synchrotron photon flux. Therefore, a quadratical relation is expected by synchrotron photon flux and VHE gamma-ray flux if the flux variation is dominated by the variation of the electron density. Such a relation is clearly obtained in [12] when systematically study the averaged flux evolution in several years period. However, the Tibet AS-$\gamma$ result seems to favor a linear correlation rather than a quadratic one during the flaring period in year 2000 and 2001 [8]. These results may suggest that the dominated physical parameters, such as magnetic field strength, electron density and its spectrum and so on, for the flux variation may be differ flare by flare as suggested in [22].

\subsection{Mrk 501}
Mrk 501 (z = 0.034)  was discovered with VHE emission by the Whipple collaboration [23]. It is also one of the  most  studied blazars. The variation of its emission  differs  from that of Mrk 421. In 1997, Mrk 501 went into a state with surprisingly high activity and strong variability and became more than a factor 10 brighter (above 1 TeV) than the Crab Nebula, and the VHE gamma-ray photons with energy up to 20 TeV is observed [24].
During this period, a marginal signal at the 3.7$\sigma$ level was observed by the Tibet AS-$\gamma$ experiment. During the most rapid increasing period, the signal was 4.7$\sigma$ [9].
In the following 14 years, the emission is steady except  of the fastest VHE gamma-ray flux variability on a timescale of minutes  observed in 2005 [25].  In October 2011, Mrk 501 underwent a strong flare in X-rays   and the emission of VHE gamma-rays is flaring as detected by the ARGO-YBJ detector [14]. Unfortunately, no IACT can observed this source since it appeared in day time.

Mrk 501 was also continuously monitored by ARGO-YBJ since 2007. The flux  was been detected by ARGO-YBJ with significance about 5$\sigma$ during the steady phase [14]. The flux entered into a high level since October 2011 and at least last to about July 2012 according to X-ray result shown in Fig.10.
The light curves from Mrk 501 at energy 15$-$50 keV is provided by $Swift$/BAT
\footnote{ Transient monitor results provided by the $Swift$/BAT team:
http://heasarc.gsfc.nasa.gov/docs/swift/ results/transients/weak/Mrk501/.} and at energy   (2$-$12 keV) is provided by   ASM/$RXTE$\footnote{Quick-look results provided by the ASM/$RXTE$ team:
\url{http://xte.mit.edu/ASM_lc.html}}.
During the first flare from October 17 to November 22, 2011, the VHE gamma-ray flux was detected by ARGO-YBJ  with significance greater than 6$\sigma$, corresponding to an factor of 6.6$\pm$2.2 times enhanced comparing to  its steady emission. In particular, the gamma-ray flux above 8 TeV is detected with a significance better than 4$\sigma$, which did not  occur  since the 1997 flare. The SED  is harder than those observed during the flares in 1997 [24] and in June 30, 2005 flare [25], and it favors the EBL model with the least absorption to VHE photons. However,  because of  large statistic error, no definite conclusion is derived. A simple one-zone SSC model can fit multiwavelength SED very well during the stead phase, while it is not able to reproduce the spectral shape at energy above 6 TeV during the flare phase, as shown in  Fig.11. Since the gamma-rays above 1 TeV are typically produced in the Klein-Nishina regime, their rate should be strongly suppressed. More details about these results  can be found elsewhere [14].

\vspace*{1mm}
\begin{center}
\centerline{\psfig{figure=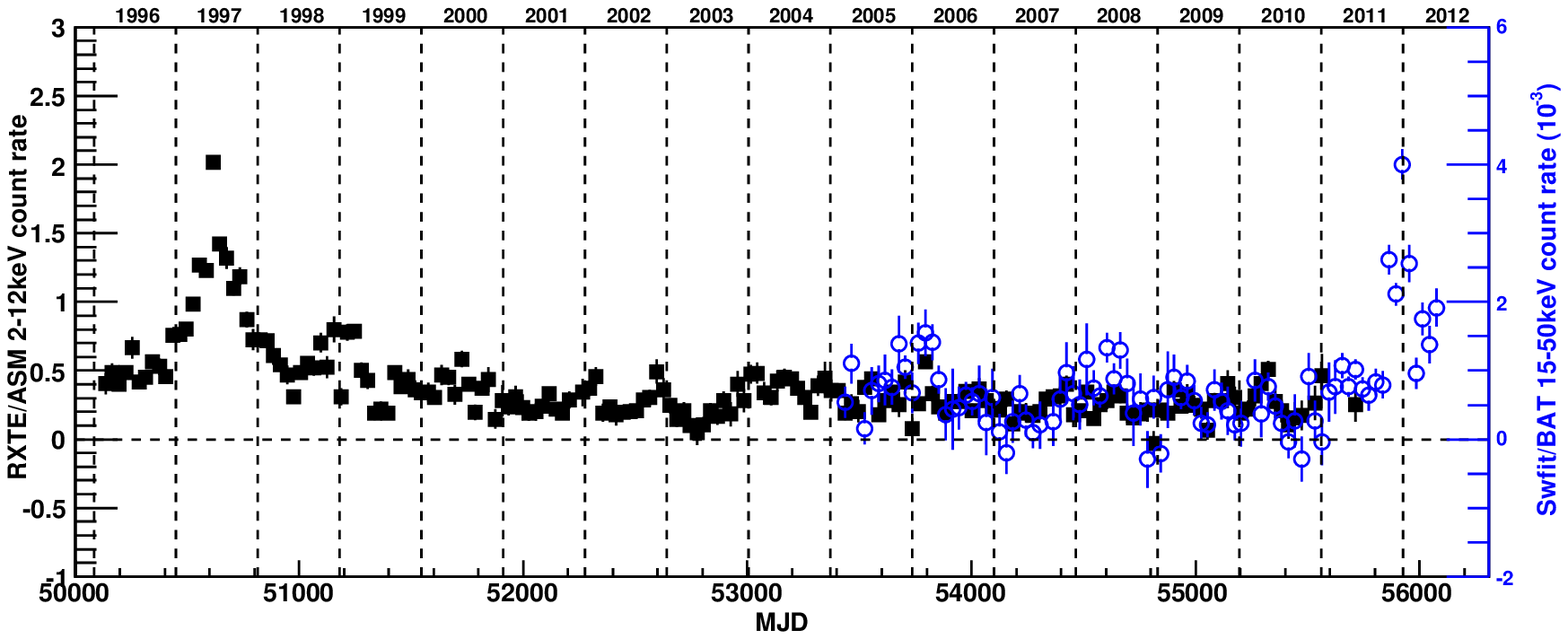,width=3.5in,height=1.5in}}
\centerline{\psfig{figure=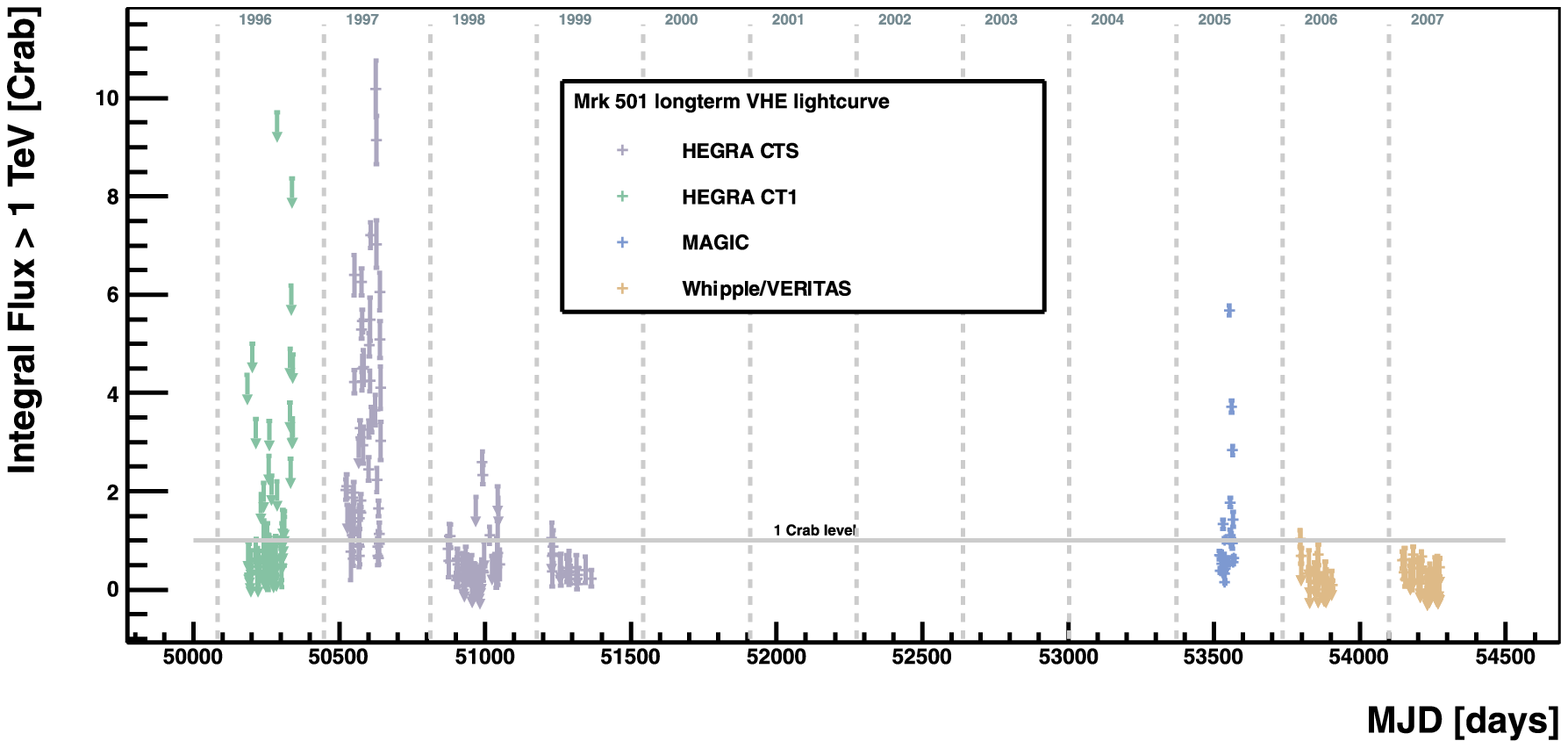,width=3.0in,height=1.5in}}
\end{center}
\vspace*{-5mm} {\footnotesize {\bf Figure 10}\quad Long-term light curve of Mrk 501. Up: Light curve observed by RXTE/ASM  at 2-12keV and Swift/BAT at 15-50keV.
Each bin contains the event rate averaged over the
30-day interval centered on that bin. Down:  A combined long-term light curve using data from  the major VHE gamma-ray telescopes, which is presented in [7]. 
Reproduced by permission of Astronomy and Astrophysics.  }

\section{Prospect of future EAS arrays}
The emission from AGNs are variable. To understand the  gamma-ray emission mechanism, long-term multiwavelength monitoring is
essential  particularly  at X-ray and VHE gamma-ray bands.  As demonstration in last section when monitor the Mrk 421 and Mrk 501, the IACT is  not optimized for long-term monitoring the VHE gamma-ray emission,  while the EAS array  is an essential  and irreplaceable approach.   However, the sensitivity of current EAS array  may  prevent us from further look insight into the properties of AGN jet. Fortunately,  two huge EAS arrays projects, HAWC and LHAASO have been proposed to be constructed in the near few years. These projects,  particularly  the LHAASO project will boost the sensitivity of current EAS array, as shown in Fig.4. These projects have a major impact on a variety of topics related to AGN physics. In the following, we will take LHAASO as an example to briefly discuss its affection for the AGN and its relative fields.

\vspace*{1mm}
\begin{center}
\centerline{\psfig{figure=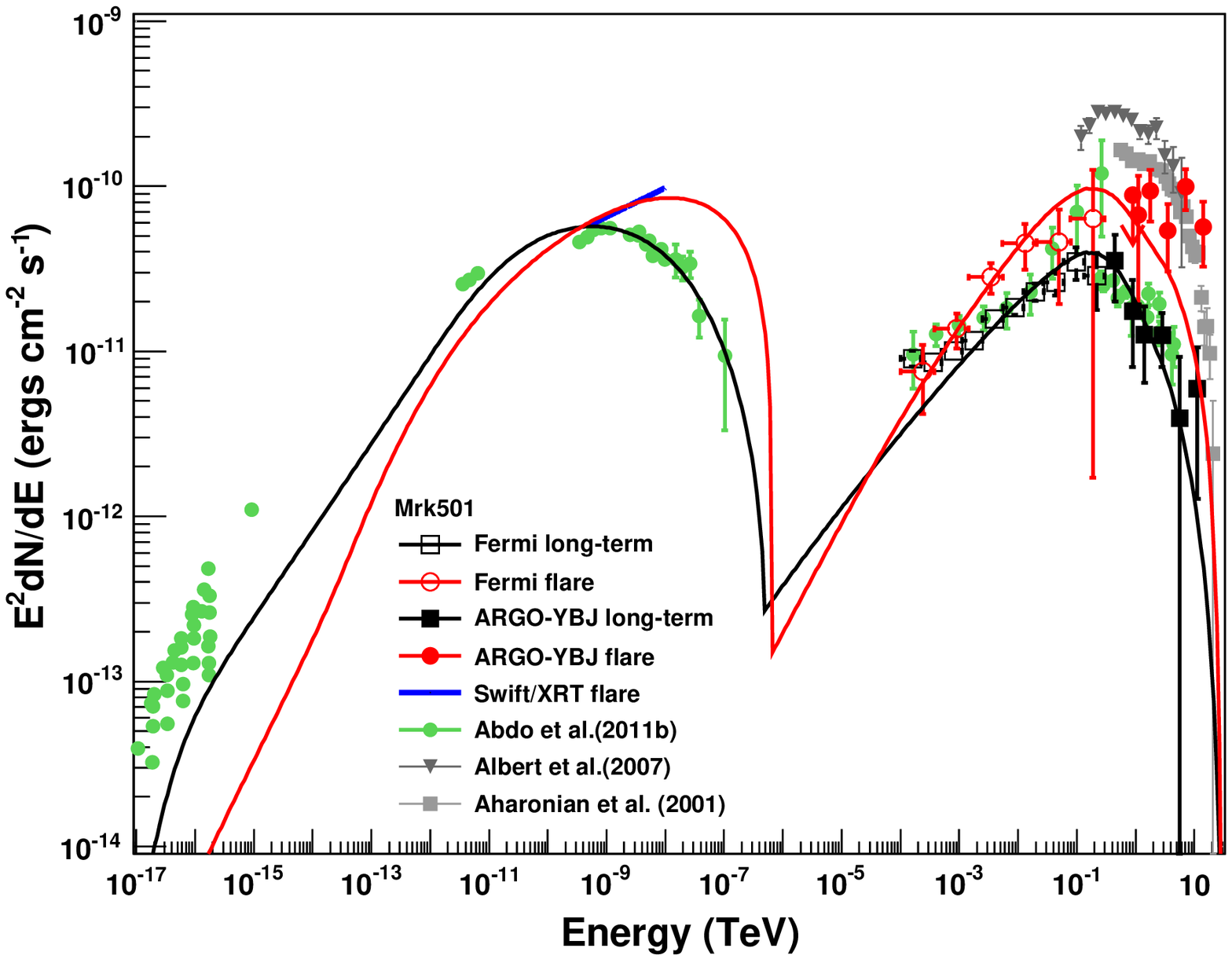,width=3.2in,height=2in}}
\end{center}
\vspace*{-5mm} {\footnotesize {\bf Figure 11}\quad Spectral energy distribution of Mrk 501.
The solid lines show the best fits to the data with a one-zone SSC model. The two curves corresponds to the SED model describing the long-term averaged
data and the flaring data, respectively. The figure is taken from [14]. Reproduced by permission of the AAS.}

\subsection{Sensitivity for different issues}
As shown in  Fig.4, the best sensitivity of LHAASO around TeV energy is 2\% Crab unit for sources located in the same declination as Crab. The sensitivity is position dependent and  varies from 2\% to 10\% when the declination of source varying from 30$^{\circ}$ to -10$^{\circ}$ and 70$^{\circ}$ according to the experience from ARGO-YBJ [26]. In the following, we will take 2\% Crab unit as a standard sensitivity for estimation, which indicates that LHAASO can achieve a 5$\sigma$ signal for source with flux of 2\% Crab unit in one year observation. Corresponding sensitivity at other declinations can be simply derived using the relations presented in [26].

 It is hoped that the sensitivity  of LHAASO will reach to   0.9\% Crab unit  in 5 years operation. This sensitivity is comparable with current IACTs, $\sim$  1\% Crab unit, with which about 50 AGNs were detected  and the number still increases with time.
LHAASO can naturally  survey  a half of the whole sky from declination -20$^{\circ}$ to 80$^{\circ}$.
Fig.12 shows the FOV of LHAASO and locations of AGN which have been detected by IACTs at VHE [2] and $Fermi$-LAT at 10-100GeV band[1]. 42 out of 53  VHE  AGNs and 271 out of 360 $>$10 GeV AGNs are inside the FOV of LHAASO, which will be possibly detected by LHAASO at VHE band.
The survey   will not only detect the known AGNs, and also enlarge number sources and probably  discover new classed of AGN, not yet known to emit at VHE, such as radio-quiet AGN. Such a survey will provide a good sample for population study, and also provide an important guide for the narrow FOV but more sensitive IACTs, which will significant improve their effective observation time.

\vspace*{4mm}
\begin{center}
\centerline{\psfig{figure=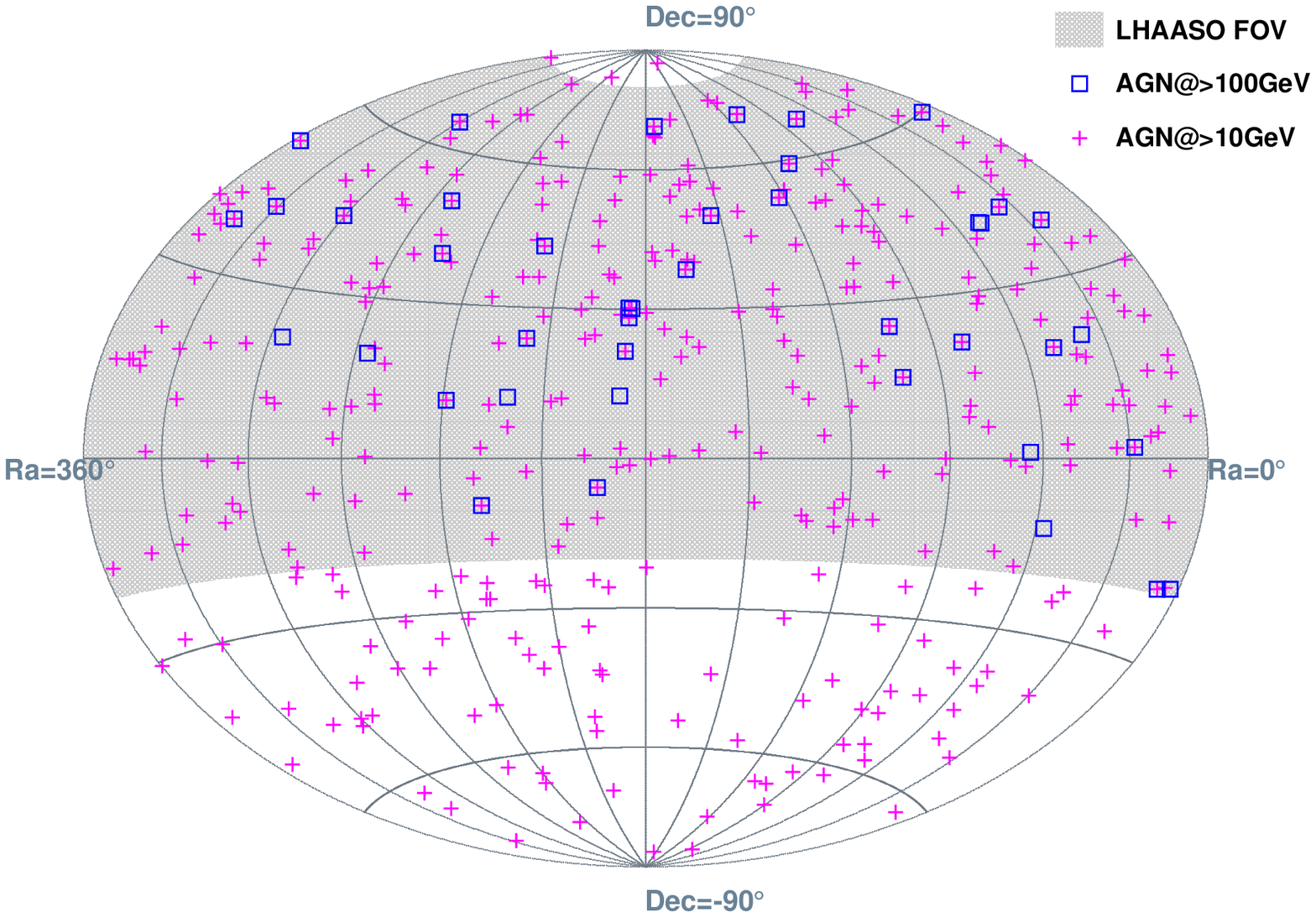,width=3.8in,height=2.7in}}
\end{center}
\vspace*{-5mm} {\footnotesize {\bf Figure 12}\quad
Distribution of discrete  AGNs with possible  gamma-ray emission at VHE band. The sky map
is in equatorial coordinates. The 53 AGNs with VHE gamma-ray emission  is taken from [2](as of December 2012).
The 360 crosses indicate the AGNs which have been significantly detected by $Fermi$-LAT at 10-100GeV band using the first two-year data[1].
The shadow region indicates the FOV of LHAASO.  
}

The sensitivity  is 38\%, 14\% and 7\% Crab unit for the observation in one day, one week and one month, respectively. With such sensitivity, the bright AGNs such as Mrk 421 and Mrk 501 can be monitoring day by day even in the quiet state. Their  flux is about 30\% Crab unit in the quiet state, which can be enhanced as a factor of tens during the flaring states.   With the long-term data of these bright sources, the solid temporal and spectral features of AGNs can be obtained, which will shed light on the underline physics of jet and its emission. As an all sky monitor, LHAASO will be an important  alert system on unexpected short flares and then conduct   multiwavelength campaign  as presented in [27].

\subsection{Long-term monitor at VHE band}
Despite the large number of blazars observed by IACTs with VHE gamma-ray emission, the
mechanisms inside these sources that are responsible for the detected emission and the underlying particle acceleration processes
are still unknown. The main reasons for these are the sparse
multifrequency data, especially at VHE band, during long periods of time.
Although the ARGO-YBJ experiment have made substantial contributions for long-term monitoring of Mrk 421 and Mrk 501,
its limited sensitivity prevent us from further looking insight into the details of temporal features and corresponding spectral evolution.
With a much better sensitivity than ARGO-YBJ, LHAASO will  have  an important role in long-term monitoring on AGN at VHE band, which is
of crucial importance to reveal the physical processes underlying  the blazar emission.

In the popular leptonic model, the gamma-ray emission from AGN is interpreted as high energy electron inverse Compton scattering of low energy pothons, however, the alternative hadronic model cannot be excluded, which can also interpret the observed SED facing no insurmountable objections based on power or energetics [28]. A main  method  to distinguish these two models is using long-term simultaneous broadband data,  particularly  at VHE gamma-ray and X-ray band which are with high variation. The ideal targets are high-frequency-peaked BL Lac objects since the gamma-ray emission is expected to be dominated by SSC with little contribution from EC [29].
In the framework of SSC model, the VHE gamma-ray and X-ray emissions are from the same electron population, therefore, their flux variability are expected to be closely  correlated. While this correlation is loose in the framework of hadronic model. ARGO-YBJ have obtain a zero day lag for the correlation study (see Fig.9), while the statistics error may be not  sufficiently small for this distinguish. With future LHAASO data, if the correlation time scale can be constrained to be within sub-hour or even shorter, it will be strong evidence to support the leptonic model. Equally important as the study of  flux correlation will be the spectral correlation since a similar spectral evolution are predicted in SSC model. Such a spectral correlation is supported by the 3-year averaged ARGO-YBJ and X-ray data at different flux levels [12]. This  correlation can be thoroughly tested in the LHAASO era using short term data in different periods.

VHE gamma-ray observations can also afford excellent opportunities to study possible particle acceleration mechanisms, diffusive shock acceleration   or magnetic reconnection.
In fact, a simultaneous broadband SED can act as a snapshot of the emitting population of particles at a given time.
High energy particles  radiative cooling time scales is   short (minute), therefore,  the VHE gamma-ray observations, comparing to longer wavelength, can perform better to track the temporal evolution of the particle energy spectra.
It will be interesting to compare the results from the SED modeling of blazars with the predictions of the theories of particle acceleration in different mechanisms. The difficulty is even in the simplest one-zone SSC model of blazar emission, the spectral evolution can be explained with different scenarios, such as the change of injected particle energy distribution, circumambient  magnetic, or Doppler factor of the emission region. To offer stronger constraints on the model parameters,  time-dependent calculations are  needed to achieve a robust intrinsic  particle energy spectra [30].
The higher sensitivity of LHAASO within a broad energy band accompanied with its long-term monitoring,  the time-dependent SED for all the flares can be well resolved from pre to post of the flares, which would permit a detailed time-dependent model taking into account the acceleration, cooling and other parameter evolution. Different scenarios may could be distinguish and thus we will be close to reveal the intrinsics process of the jet.

Variability is frequently detected in blazars. The time scales of variability vary in a large range from days to months, especially, rapid flare down to minutes has been detected recently.
For rapid flare, both X-ray and gamma-ray ``orphan" flares were reported  [21], which are taken as the challenge to the SSC models. However, because of  the limitation of temporal coverage, these observations can not rule out the possibility of delay between the two bands.
Well measured  long-term multiwavelength  light curves  are  critical  to investigate the features of the variability, which are crucial to reveal the underlying physics processes.  This may include  whether the flares are general features at all flux states or are added features separated to the static quiescent states. It would also be helpful to reveal the relation between rapid flares (referred to sub-hour variability) and the general flares (with variable timescale from day to month), whether they are belong to the same phenomenon with random time scale or are attached to  different catalogs with different underline mechanism.  In the LHAASO era, these long-term variability will be well investigated with a population of well determined light curves from different AGNs. Such a study will provide a global picture of the evolution of the underlying particle energy spectra, which will be crucial importance for the study of emission mechanisms and particle acceleration mechanisms inside the jet.

\subsection{ Observation of gamma-ray emission  from Radio Galaxies}
Unlike the blazars, the jet of radio galaxies,  is not directly oriented towards the line-of-sight, which provide an alternative
method  to investigate the location of the gamma-ray emission region.
Given the typical angular resolution of IACTs ($\sim$ 0.1$^{\circ}$)  and EAS arrays ($\sim$ 0.5$^{\circ}$) , it is  not possible  to directly determine the location of the VHE emission region.
However, with the help from radio to X-ray wavelengths which can resolve the jets  structure with much better angular resolution (better than 1 milli-arc sec),  the location of the VHE emission region can be determined. The case of M87
demonstrates effectiveness of the approach during the VHE flaring periods in 2005, 2008 and 2010 [31-33]. The 2008 multiwavelength observations
revealed the VHE gamma-ray flare accompanied
by a strong increase of the radio flux from its nucleus, which imply that charged particles
are accelerated to very high energies in the immediate vicinity of the black hole [32]. The 2010 flare shows similar timescales and peak flux
levels as that in 2005 and 2008 at VHE band, while the multiwavelength  variability
pattern  appears somewhat difference  and no enhanced flux at radio band is observed [33]. To fully figure out the picture of such flare phenomena, we need collect more events in the future.

The crucial of the approach is to achieve a long-term monitoring observation at multiwavelength and to search for the correlation gamma-ray emission with the changes in the jet structure. Obviously, the EAS array is a ideal approach for long-term monitoring at VHE gamma-ray band as discussed in previously. Currently, three radio galaxies have been identified as VHE emitter, i.e., Centaurus A, M87 and NGC 1275. The flux M87 has reached 20\% Crab units during the flaring period [32,33], which can be detected by LHAASO within one week observation. With the observation of LHAASO, the number of flaring events from M87 will increase  and thus provide more evidence to identify the location of VHE gamma-ray emission. Another contribution from LHAASO for this issue is that new radio AGNs may be discovered with sky survey and provide new targets to IACTs for deep observation.
It can be concluded that  the originates of VHE emission  will be well located  with collective data samples in the LHAASO era.

High energy gamma-ray extended emission from the lobes of Centaurus A has been discovered with the $Fermi$-LAT and the angular extension of the lobes is $\sim10^{\circ}$ [34].
The gamma-ray emission from the lobes detected
by the LAT is interpreted as inverse Compton scattering the cosmic microwave background (CMB), with the infrared-to-optical extragalactic back-ground light  contributing at higher energies.
The $Fermi$-LAT data  cannot distinguish wether the electron are
accelerated in situ or efficiently transported from
regions closer to the nucleus [34]. If VHE gamma-ray emission is observed from lobes region, it will be strong evidence for the particle acceleration in the lobes since  their radiative lifetimes are too short to be transported through the few
hundred kiloparsec-scale lobes.
At VHE gamma-ray band, such a large extend region is  better observed by wide FOV EAS arrays. Obviously, Centaurus A will not be inside LHAASO FOV, however, search new similar emission in northern sky  can be done.
\subsection{Probing the EBL with VHE gamma-rays}
On their way from the AGNs to us, a fraction of the gamma-rays are absorbed by the extragalactic
background photon, due to electron-positron  pair production $\gamma_{VHE}+\gamma_{EBL}\rightarrow e^{-} + e^{+}$.
In general, for VHE photons the domination is the optical/infrared background
radiation, which is called extragalactic background light (EBL).  Conversely,  VHE spectra can thus be used to probe the EBL density, which remained an observational challenge for direct measurements while
contains important information both the evolution of baryonic components of galaxies and the structure of the Universe in the pre-galactic era.
Upper limits on the EBL have been derived by IACTs using different distant blazars, such as H2356-309, 1ES 1101-232, 1ES0229+200 and 3C 279 [35-37],  under the assumption that the intrinsic photon index of the source is not harder than 1.5 [38]. These limits are  close to the lower limit derived from direct measurements of the integrate light of resolved galaxies.  Combining the high energy ($Fermi$) and VHE spectra,  ARGO-YBJ collaboration recently also provide a similar conclusion typically using gamma-rays around 10 TeV basing the very closed blazar Mrk 501 [14], when it is suffering a large flare and its VHE spectrum is hardest than anyone previous. Comparing to future IACT array CTA, LHAASO is not so sensitive for specific source especially for distant sources.  However, LHAASO still have its superiority at least in two respects,  that is population study and capture flares with hard spectrum. LHAASO will survey half of the whole sky, therefore, a large mount of AGNs can be detected and thus be used to give a statistical  result on  the EBL density.  Because of its high duty cycle, LHAASO has much more possibility than IACTs to capture flares with hard spectrum like that presented in Mrk 501.

\subsection{Estimate the Intergalactic magnetic fields (IGMF) with VHE gamma-rays}
The electron-positron pair produced by VHE gamma-ray with EBL can emission high energy gamma-rays  because of  Inverse Compton up-scattering the cosmological microwave background photons. This can result in detectable signatures which  depends on the strength of the IGMF.
If the IGMF is weaker than $10^{-16}$G, delayed high energy gamma-ray signal from the direction of the source could be detected after the VHE gamma-ray flares. This phenomenon is called ``pair-echos" [39].   For stronger magnetic field (such as $\sim 10^{-14}$G ), the same process can lead to extended 0.1-10 degree scale emission at multi-GeV and TeV energies around the source [40], which is called ``pair-halos" and can be detected in the VHE range. In order to detect the ``pair-echos", a  long-term simultaneous GeV-TeV light curves is crucial. For the ``pair-halos", a wide-FOV VHE detector is better since the extension of the halo could be several degrees.  The typical targets for this issue are bright blazars, such as Mkr 421 and Mrk 501. Therefore, the LHAASO is  suitable   because of  its high duty cycle, large FOV and excellent sensitivity to the targets.
 The energy resolution of LHAASO at the sub-TeV energy band is about 30\%, which is  sufficient  to pick out the low energy gamma-rays for ``pair-halos" study. The angular resolution is around 0.7$^{\circ}$ with which   the extended ``pair-halos" emission can be distinguished from point source.
A good demonstration for such situation is the case of ARGO-YBJ, which share a comparable
angular resolution  and have determined the extension sizes of three  sources to be (0.2$\sim$0.5$^{\circ}$) [41-43].

\subsection{Lorentz Invariance Test }
During the rapid flares from AGN, the observation of gamma-rays can also be used to set limits on Lorentz invariance violation (LIV) [44], which is predicted by certain quantum gravity theories. The induced time delay of the dispersion becomes larger with increasing photon energy and distance of the source. The sensitivity of LHAASO is not  as  good as the CTA in short-time observation. However, LHAASO has much more opportunity to catch large flare. Probability, it may achieve a good result on this issue with help of large flares and its wide energy range.

\section{Conclusion}
In the past decade,  about 50 AGNs were detected by IACTs with VHE gamma-ray emission and their  fluxes were frequently detected with strong variability. To further understand to emission mechanism and the intrinsic process inside the jet of AGNs, beside to enlarge the number of AGN sample, long-term monitor data is also essential to obtain a full picture of the variability.  Because of  the limitation of IACTs, which
cannot operate during non-optimal weather conditions or bright moonlight periods, EAS array is irreplaceable to achieve a long-term monitoring at VHE.
The EAS arrays Tibet AS-$\gamma$ and ARGO-YBJ experiments have demonstrated the superiority of their high duty cycle when monitor the Mrk 421 and Mrk 501 even with limited sensitivity. The planing future projects, such as LHAASO, will  boost the sensitivity up to 30 times.
With such excellent sensitivity, a long-term detailed variability of the emission from AGNs will be revealed including both the flux and the spectral evolution. These observation data will have a major impact on our knowledge of the process inside the AGN jet, such as the origin and physics mechanism
responsible for the VHE emission in AGNs and thereby probe the conditions of particle acceleration and
cooling in relativistic plasma outflows and in the vicinity of super-massive black holes. Important implication are also expected for related field such as the density of extragalactic background light, the strength of intergalactic magnetic fields, and the validity of the Lorentz Invariance. Additionally, an important role of EAS array will   be   sky monitor and  to send alert to IACTs for deep observation, which have an excellent sensitivity for short variation.

\Acknowledgements{\bahao
 Project  11205165   supported by National Natural Science Foundation of China. This work is also supported by Xiejialin Fund (Y3546140U2) of IHEP, CAS.
}


\normalsize \vskip0.3in\parskip=0mm \baselineskip 18pt
\renewcommand{\baselinestretch}{1.1}\footnotesize\parindent=4mm\bahao

\REF{1\ }  Nolan P L,    Abdo  A  A,   Ackermann M,  et al. Fermi large area telescope second source catalog. Astrophys J Suppl, 2012, 199:31
\REF{2\ } Aharonian  F,   Buckley J,  Kifune T,  et al. High energy astrophysics with ground-based gamma ray detectors.   Rep  Prog Phys, 2008, 71:096901 
\REF{3\ } Dermer  C  D, Schlickeiser  R,   Mastichiadis A. High-energy gamma radiation from extragalactic radio sources. Astron Astrophys, 1992, 256:L27$-$L30
\REF{4\ } Ghisellini  G,   Celotti  A, Fossati G, et al.  A theoretical unifying scheme for gamma-ray bright blazars. MNRAS, 1998, 301:451$-$468
\REF{5\ } Aharonian  F. TeV gamma rays from BL Lac objects due to synchrotron radiation of extremely high energy protons.  New Astron, 2000, 5:377$-$395
\REF{6\ } Aharonian  F, Akhperjanian  A  G, Bazer-Bachi  A  R, et al. An exceptional very high energy gamma-Ray flare of PKS 2155$-$304.  Astrophys J, 2007,  664:L71$-$L74
\REF{7\ } Tluczykont  M,  Bernardini E,   Satalecka K, et al. Long-term light curves from combined unified very high energy gamma-ray data. Astron Astrophys, 2010, 524:48
\REF{8\ } Amenomori  M,  Ayabe S,   Cui S W, et al. Multi-TeV Gamma-Ray Flares from Markarian 421 in 2000 and 2001 Observed with the Tibet Air Shower Array. Astrophys J, 2003, 598:242$-$249
\REF{9\ } Amenomori M, Ayabe S, Cao P Y, et al. Detection of multi-TeV gamma rays from Markarian 501 during an unforeseen flaring state in 1997 with the Tibet air shower array. Astrophys J, 2000, 532:302$-$307
\REF{10\ } Atkins  T,  Benbow W,  Berley D,  et al. TeV gamma-ray survey of the northern hemisphere sky using the Milagro observatory. Astrophys J, 2004, 608:680$-$685
\REF{11\ } Aielli  G, Assiro R, Bacci  C, et al. Layout and performance of RPCs used in the Argo-YBJ experiment. Nucl  Instrum  Methods Phys Res  A, 2006, 562:92$-$96
\REF{12\ } Bartoli  B, Bernardini  P, Bi  X  J, et al. 	Long-term Monitoring of the TeV Emission from Mrk 421 with the ARGO-YBJ Experiment. Astrophys J, 2011, 734:110
\REF{13\ } Aielli G, Assiro R, Bacci  C, et al. Gamma-ray Flares from Mrk421 in 2008 Observed with the ARGO-YBJ Detector.  Astrophys J  Lett, 2010,  714: L208$-$L212
\REF{14\ } Bartoli  B, Bernardini  P, Bi  X  J, et al. Long-term Monitoring on Mrk 501 for Its VHE gamma Emission and a Flare in October 2011. Astrophys J, 2012, 758:2
\REF{15\ } Angelis A,   Mansutti O, Persic M. Very-high energy gamma astrophysics. arXiv:07120315
\REF{16\ } He  H H. LHAASO Project: detector design and prototype. Proc  31st ICRC, LORZ, Poland, 2009, (\url{http://icrc2009.uni.lodz.pl/proc/pdf/icrc0654.pdf})
\REF{17\ } Punch  M,   Akerlof C W,  Cawley M F,  et al. Detection of TeV photons from the active galaxy Markarian 421.  Nature, 1992, 358:477$-$478
\REF{18\ } Wagner  R. Correlated variability in Blazars. 2008, arXiv:0808.2483
\REF{19\ } He  H H, Chen S Z,  Zhang L. A multi-wavelength view of the large gamma-ray flares from Mrk 421 in 2010. Proc  32nd ICRC, Beijing, China, 2011, 8:181, (\url{http://www.ihep.ac.cn/english/conference/icrc2011/paper/proc/v8/v8_1209.pdf})
\REF{20\ } Chen S Z, Long-term monitor on Mrk 421 TeV emission using ARGO-YBJ experiment. Proc  32nd ICRC, Beijing, China, 2011, 8:141, (\url{http://www.ihep.ac.cn/english/conference/icrc2011/paper/proc/v8/v8_1007.pdf})
\REF{21\ } Blazejowski  M,  Blaylock G,   Bond I H, et al. A multiwavelength view of the TeV Blazar Markarian 421: correlated variability, flaring, and spectral evolution. Astrophys J, 2005, 630:130$-$141
\REF{22\ } Katarzynski  K, Ghisellini  G, Tavecchio F, et al. Correlation between the TeV and X-ray emission in high-energy peaked BL Lac objects. Astron Astrophys, 2005, 433:479$-$496
\REF{23\ } Quinn  J, Akerlof C W, Biller S, et al. Detection of Gamma Rays with E $>$ 300 GeV from Markarian 501. Astrophys J Lett, 1996,  456:L83
\REF{24\ } Aharonian F, Akhperjanian  A G, Barrio J A, et al. The temporal characteristics of the TeV gamma-radiation from Mrk 501 in 1997. Astron Astrophys, 1999, 342:69$-$86
\REF{25\ } Albert  J, Aliu E, Anderhub H, et al. Variable Very High Energy gamma-Ray Emission from Markarian 501. Astrophys J, 2007,  669:862$-$883
\REF{26\ } Cao  Z, Chen  S  Z.  TeV gamma-ray survey of the northern sky using the ARGO-YBJ experiment.  Proc  32nd ICRC, Beijing, China, 2011, 7:212, (arXiv:1110.1809)
\REF{27\ } Bartoli  B, Bernardini  P, Bi  X  J, et al. Early warning for VHE gamma-ray flares with the ARGO-YBJ detector. Nucl  Instr  Meth A, 2011, 659:428$-$433
\REF{28\ } Abdo  A A,  Ackermann M,  Ajello M, et al. Fermi-LAT observations of Markarian 421: the missing piece of its spectral energy distribution. Astrophys J, 2011, 736:131
\REF{29\ } Eileen T M, Giovanni F, Markos G, et al. Collective Evidence for Inverse Compton Emission from External Photons in High-Power Blazars. Astrophys J Lett, 2012, 752:L4
\REF{30\ } Li  H,  Kusunose  M. Temporal and spectral variabilities of high-energy emission from blazars using synchrotron self-compton models.   Astrophys J, 2000, 536:729$-$741
\REF{31\ } Aharonian  F, Akhperjanian  A  G, Bazer-Bachi  A  R, et al. Fast variability of Tera-electron volt gamma-rays from the radio galaxy M87.     Science, 2006, 314: 1424$-$1427
\REF{32\ } Acciari  V A,   Aliu E,  Arlen T, et al. Radio imaging of the Very-High-Energy gamma-ray emission region in the central engine of a radio galaxy.  Science, 2009, 324:444$-$448
\REF{33\ } Abramowski  A,   Acero F,   Aharonian F, et al. The 2010 very high energy gamma-ray flare and 10 years of multi-wavelength observations of M 87.  Astrophys J, 2012, 746:151
\REF{34\ } Abdo  A A,   Ackermann M,  Ajello M, et al. Fermi Gamma-Ray Imaging of a Radio Galaxy. Science, 2010, 328:725$-$729
\REF{35\ } Aharonian F, Akhperjanian  A  G, Bazer-Bachi  A  R, et al. A low level of extragalactic background light as revealed by gamma-rays from blazars.   Nature, 2006, 440: 1018$-$1021
\REF{36\ } Aharonian F, Akhperjanian  A  G,   Almeida U B, et al. New constraints on the Mid-IR EBL from the HESS discovery of VHE gamma rays from 1ES 0229+200.  Astron Astrophys, 2007, 475:L9$-$L13
\REF{37\ } Albert  J,   Aliu E,  Anderhub H, et al. Very high energy gamma rays from a distant Quasar: How transparent is the Universe.  Science, 2008, 320:1752$-$1754
\REF{38\ } Malkov  M  A, O'C Drury L. Nonlinear theory of diffusive acceleration of particles by shock waves. Reports on Progress in Physics, 2001, 64:429$-$481
\REF{39\ } Inoue  S, Takahashi K, Mori M, et al. Probing Intergalactic Magnetic Fields with Gamma Rays from Blazars. PoS (AGN 2011) 032
\REF{40\ } Elyiv  A, Neronov A, Semikoz D.  Gamma-ray induced cascades and magnetic fields in the intergalactic medium. Physical Review D, 2009, 80:23010
\REF{41\ } Bartoli  B, Bernardini  P, Bi  X  J, et al.  Observation of TeV Gamma Rays from the Cygnus Region with the ARGO-YBJ Experiment. Astrophys J Lett, 2012, 745:L22
\REF{42\ } Bartoli B, Bernardini  P, Bi  X  J, et al. Observation of the TeV gamma-ray source MGRO J1908+06 with ARGO-YBJ.  Astrophys J, 2012, 760: 110
\REF{43\ } Bartoli B, Bernardini  P, Bi  X  J, et al. Observation of the TeV gamma-ray from the unidentified source HESS J1841-055 with the ARGO-YBJ experiment.  Astrophys J, 2013, 767: 99
\REF{44\ } Abramowski  A,   Acero F,   Aharonian  F, et al. Search for Lorentz Invariance breaking with a likelihood fit of the PKS 2155-304 Flare Data Taken on MJD 53944. Astropart  Phys, 2011, 34:738$-$747

\end{multicols}

\end{document}